\pdfoutput=1

\documentclass[11pt]{article}

\usepackage{acl}

\usepackage{times}
\usepackage{latexsym}

\usepackage[T1]{fontenc}

\usepackage[utf8]{inputenc}

\usepackage{microtype}

\usepackage{inconsolata}

\usepackage{graphicx}
\usepackage{amsmath}
\usepackage{amsfonts}
\usepackage{booktabs}
\usepackage{multirow}
\usepackage{subcaption}
\usepackage{import}
\usepackage{lipsum}
\usepackage[linesnumbered, boxed]{algorithm2e}
\usepackage{makecell}
\usepackage{amssymb}
\usepackage{fontawesome}
\usepackage[table,dvipsnames]{xcolor}
\newcommand{\review}[1]{\textcolor{black}{#1}}
\usepackage{blindtext}
\usepackage{url}
\usepackage{tabularray}
\usepackage{tablefootnote}

\usepackage{tcolorbox}
\usepackage{spverbatim}
\usepackage{xcolor}
\usepackage{fancyvrb}
\usepackage{fvextra}
\usepackage{array}
\usepackage[flushmargin]{footmisc}

\title{T$^2$-RAGBench: Text-and-Table Benchmark for Evaluating \\ Retrieval-Augmented Generation}

\author{
 \textbf{Jan Strich\textsuperscript{1}},
 \textbf{Enes Kutay Isgorur\textsuperscript{2}},
 \textbf{Maximilian Trescher\textsuperscript{2}}, \\
 \textbf{Chris Biemann\textsuperscript{1}},
 \textbf{Martin Semmann\textsuperscript{1}} \\
 \\
 \textsuperscript{1}University of Hamburg, Germany \\
 \textsuperscript{2}dida Datenschmiede GmbH \\
 \\
 \small{
   \textbf{Correspondence:} \href{mailto:t2ragbench@gmail.com}{t2ragbench@gmail.com}
 }
}

\begin{document}
\maketitle

\begin{abstract}
Since many real-world documents combine textual and tabular data, robust Retrieval Augmented Generation (RAG) systems are essential for effectively accessing and analyzing such content to support complex reasoning tasks.
Therefore, this paper introduces \textbf{T$^2$-RAGBench}, a benchmark comprising \textbf{23,088} question-context-answer triples, designed to evaluate RAG methods on real-world text-and-table data.
Unlike typical QA datasets that operate under \textit{Oracle Context} settings, T$^2$-RAGBench challenges models to first retrieve the correct context before conducting numerical reasoning. Existing QA datasets containing text-and-table data typically contain context-dependent questions, which may yield multiple correct answers depending on the provided context.
To address this, we transform SOTA datasets into a context-independent format, validated by experts as 91.3\% context-independent questions, enabling reliable RAG evaluation.
Our comprehensive evaluation identifies \textit{Hybrid BM25}, a technique that combines dense and sparse vectors, as the most effective approach for text-and-table data. However, results demonstrate that T$^2$-RAGBench remains challenging even for SOTA LLMs and RAG methods. Further ablation studies examine the impact of embedding models and corpus size on retrieval performance.
T$^2$-RAGBench provides a realistic and rigorous benchmark for existing RAG methods on text-and-table data. Code and dataset are available online\footnote{\label{repo}\href{https://github.com/uhh-hcds/g4kmu-paper}{github.com/uhh-hcds/g4kmu-paper}}.
\end{abstract}

\section{Introduction}\label{sec:intro}

Documents containing a mixture of text and tables are widely utilized in various fields, such as financial reporting \cite{baviskar_efficient_2021}, scientific research \cite{pramanick_spiqa_2024}, and organizational documentation \cite{rebmanjr_industry_2023}.

\begin{figure}[ht]
    \centering
    \includegraphics[width=0.48\textwidth]{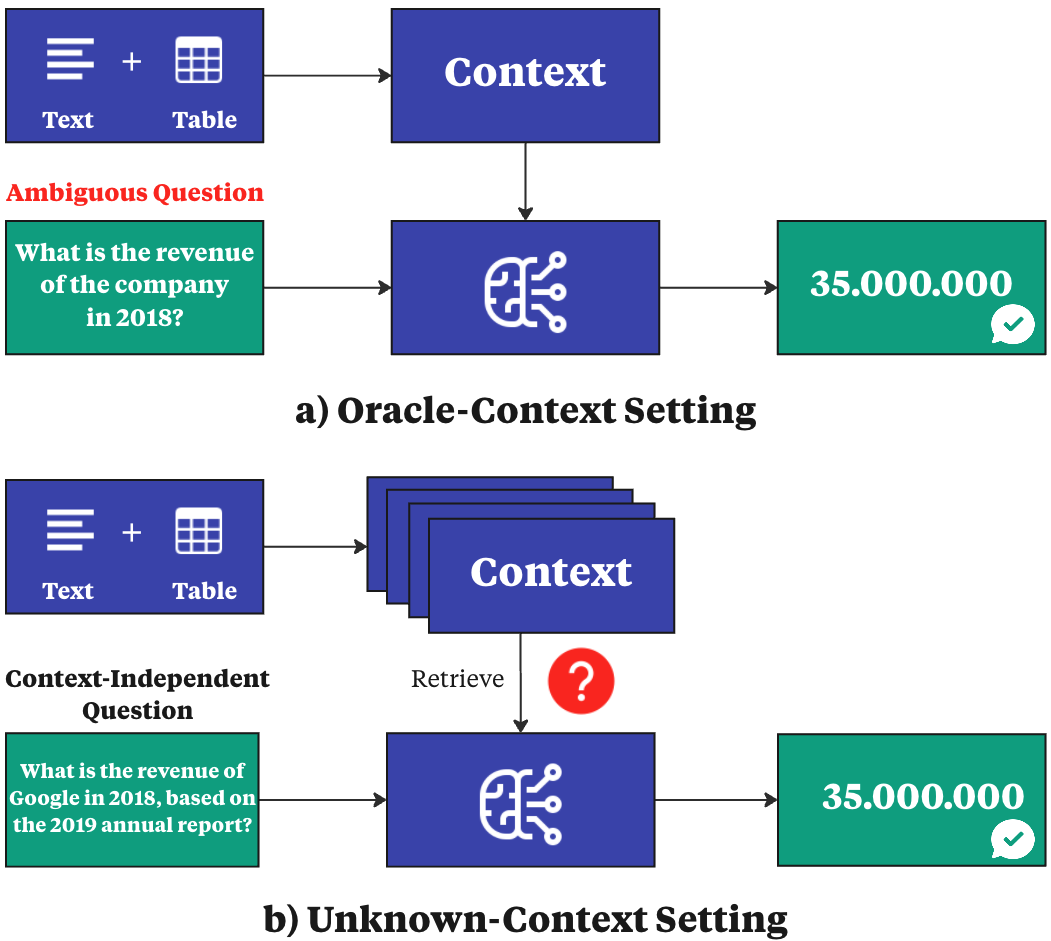}
    \caption[overview]{Overview of current SOTA approaches and dataset example. a) Most benchmarks test models in an \textit{Oracle Context} setting, \cite{chen_finqa_2021, chen_convfinqa_2022}. Our task (b) targets the unknown-context setting, requiring retrieval from mixed text-tables before answering.}
    \label{fig:overview}
    \vspace{-1.5em}
\end{figure}

Recent advancements in Large Language Models (LLMs) have demonstrated solid SOTA performance answering numerical and free-form question-answering (QA) tasks when appropriate documents are provided \cite{nan_fetaqa_2021, chen_finqa_2021, chen_convfinqa_2022, zhu_tatqa_2021, zhu_complex_2022}.
Despite increasing context window sizes for LLMs, using the entire corpus remains impractical due to computational constraints and programmatic latency \cite{wang_limits_2024a, li_long_2024}. 
Therefore, retrieving relevant documents is essential in real-world applications to answer questions correctly.

Retrieval-Augmented Generation (RAG) \cite{lewis_retrievalaugmented_2020} has emerged as a promising solution for single-hop QA on numerical tasks, providing appropriate context and has led to an explosion of methods in this area \cite{gao_retrievalaugmented_2023, nikishina_creating_2025}. While most RAG methods are effective at retrieving semantically similar text, embedding tabular data remains challenging due to its structural complexity and the predominance of numerical values, which lack semantic context \cite{khattab_demonstratesearchpredict_2022}.

In addition, RAG methods are typically trained and evaluated on text-only datasets~\cite{jiang_active_2023, lan_copy_2023, wang_limits_2024a}, Wikipedia-derived QA benchmarks~\cite{pasupat_compositional_2015, yang_hotpotqa_2018} heavily used during LLM pre-training~\cite{grattafiori_llama_2024}, or narrow domain-specific datasets~\cite{sarthi_raptor_2024, yan_corrective_2024}, making it difficult to estimate the performance on text-and-table data.
Moreover, as illustrated in Figure~\ref{fig:overview}, existing datasets with text-and-table data operate exclusively under the oracle context setting, where questions are tightly coupled with the given context. These questions are inherently ambiguous and may yield multiple correct answers depending on the context, we refer to them as \textbf{context-dependent}.
In contrast, \textbf{context-independent} questions have a single correct answer without having access to the context, which is essential for evaluating RAG methods, as they require identifying one ground truth document containing the answer. To our knowledge, no text-and-table dataset meets this requirement.

To fill this gap, we present the \textbf{T}ext-\textbf{T}able \textbf{R}etrieval-\textbf{A}ugmented \textbf{G}eneration \textbf{Bench}mark (T$^2$-RAGBench), a benchmark designed to evaluate existing RAG methods on text-table retrieval and numerical reasoning tasks. Our benchmark comprises three subsets extracted from existing datasets, totaling 23,088 question-context-answer (QCA) triples and 7,318 real-world financial documents. Each triplet includes a reformulated, context-independent question, a verified answer, and the associated context containing all information to answer the question. \\
\noindent\textbf{Our contributions are as follows:}

\begin{itemize}
    \item We introduce \textbf{T$^2$-RAGBench}, a benchmark containing \textbf{23,088} QCA triples from financial reports designed to evaluate RAG methods on text-and-table and numerical reasoning.
    \item We systematically evaluate popular RAG methods on \textbf{T$^2$-RAGBench}, demonstrating that it remains a challenging and relevant benchmark for current methods.
    \item We compare SOTA closed and open-source embedding models and analyze the effect of corpus size on promising RAG methods.
\end{itemize}

\section{Related Work}\label{sec:related_work}

\begin{table*}[ht]
\centering
\resizebox{\textwidth}{!}{%
\begin{tabular}{lcccccccc}
\toprule
\multirow{2}{*}{\textbf{Dataset}} &
\multirow{2}{*}{\textbf{Domain}} &
\multirow{2}{*}{\textbf{Text}} &
\multirow{2}{*}{\textbf{Table}} &
\textbf{Visual} & \textbf{Context-} &
\multirow{2}{*}{\textbf{Available}} &
\multirow{2}{*}{\textbf{QA Pairs}} \\
& & & & \textbf{Independence} & \textbf{Independent} & & \\
\midrule
TriviaQA~\cite{joshi_triviaqa_2017}     & Wikipedia       & \textcolor{Green}{\checkmark} & \textcolor{Mahogany}{\faTimes} & \textcolor{Green}{\checkmark} & \textcolor{Green}{\checkmark} & \textcolor{Green}{\checkmark} & 650K \\
HybridQA~\cite{chen_hybridqa_2020}      & Wikipedia       & \textcolor{Mahogany}{\faTimes} & \textcolor{Green}{\checkmark} & \textcolor{Green}{\checkmark} & \textcolor{Green}{\checkmark} & \textcolor{Green}{\checkmark} & 70K \\
FeTaQA~\cite{nan_fetaqa_2021}           & Wikipedia       & \textcolor{Mahogany}{\faTimes} & \textcolor{Green}{\checkmark} & \textcolor{Green}{\checkmark} & \textcolor{Green}{\checkmark} & \textcolor{Green}{\checkmark} & 10K \\
\midrule
Qasper~\cite{dasigi_dataset_2021}       & NLP Papers      & \textcolor{Mahogany}{\faTimes} & \textcolor{Green}{\checkmark} & \textcolor{Green}{\checkmark} & \textcolor{Mahogany}{\faTimes} & \textcolor{Green}{\checkmark} & 5K \\
SPIQA~\cite{pramanick_spiqa_2024}       & NLP Papers      & \textcolor{Mahogany}{\faTimes} & \textcolor{Green}{\checkmark} & \textcolor{Mahogany}{\faTimes} & \textcolor{Mahogany}{\faTimes} & \textcolor{Green}{\checkmark} & 270K \\
\midrule
FinQA~\cite{chen_finqa_2021}            & Finance         & \textcolor{Green}{\checkmark} & \textcolor{Green}{\checkmark} & \textcolor{Green}{\checkmark} & \textcolor{Mahogany}{\faTimes} & \textcolor{Green}{\checkmark} & 8K \\
ConvFinQA~\cite{chen_convfinqa_2022}    & Finance         & \textcolor{Green}{\checkmark} & \textcolor{Green}{\checkmark} & \textcolor{Green}{\checkmark} & \textcolor{Mahogany}{\faTimes} & \textcolor{Green}{\checkmark} & 14K \\
TAT-DQA~\cite{zhu_complex_2022}         & Finance         & \textcolor{Green}{\checkmark} & \textcolor{Green}{\checkmark} & \textcolor{Green}{\checkmark} & \textcolor{Mahogany}{\faTimes} & \textcolor{Green}{\checkmark} & 16k \\
VQAonBD~\cite{raja_icdar_2023}          & Finance         & \textcolor{Mahogany}{\faTimes} & \textcolor{Green}{\checkmark} & \textcolor{Mahogany}{\faTimes} & \textcolor{Mahogany}{\faTimes} & \textcolor{Green}{\checkmark} & 1,531K \\
FinDER~\cite{choi_finder_2025}          & Finance         & \textcolor{Green}{\checkmark} & \textcolor{Green}{\checkmark} & \textcolor{Green}{\checkmark} & \textcolor{Green}{\checkmark} & \textcolor{Mahogany}{\faTimes} & 50K \\
\midrule
DocVQA~\cite{tito_document_2021}        & Multiple        & \textcolor{Mahogany}{\faTimes} & \textcolor{Green}{\checkmark} & \textcolor{Mahogany}{\faTimes} & \textcolor{Mahogany}{\faTimes} & \textcolor{Green}{\checkmark} & 50K \\
TableBench~\cite{wu_tablebench_2025}    & Multiple        & \textcolor{Green}{\checkmark} & \textcolor{Green}{\checkmark} & \textcolor{Mahogany}{\faTimes} & \textcolor{Mahogany}{\faTimes} & \textcolor{Green}{\checkmark} & $\sim$1K \\
UDA~\cite{hui_uda_2024}                 & Multiple        & \textcolor{Green}{\checkmark} & \textcolor{Green}{\checkmark} & \textcolor{Green}{\checkmark} & \textcolor{Mahogany}{\faTimes} & \textcolor{Green}{\checkmark} & 30K \\
\midrule
\textbf{T$^2$-RAGBench (Ours)}          & Finance         & \textcolor{Green}{\checkmark} & \textcolor{Green}{\checkmark} & \textcolor{Green}{\checkmark} & \textcolor{Green}{\checkmark} & \textcolor{Green}{\checkmark} & 23K \\
\bottomrule
\end{tabular}
}
\caption{Summary and comparison of Q\&A datasets. Visual Independence: The contexts are presented as text and are not only images. Context-Independent: Without a context, questions still only have one unambiguous answer. }
\label{tab:dataset_comparison}
\end{table*}

\paragraph{Text-and-Table QA Datasets.}

Table \ref{tab:dataset_comparison} gives an overview of existing Q\&A datasets containing text and/or tables.
While datasets in common knowledge \cite{joshi_triviaqa_2017, chen_hybridqa_2020, nan_fetaqa_2021}, scientific documents~\cite{pramanick_spiqa_2024, dasigi_dataset_2021}, or medicine~\cite{fan_medodyssey_2025} focusing exclusively on tables \cite{katsis_aitqa_2022}, combining text with tables becomes essential for effectively parsing whole PDF documents. 
Another challenge is data contamination, as common knowledge and scientific datasets often rely on Wikipedia or open-access papers, which are heavily used during LLM pretraining \cite{grattafiori_llama_2024}. This makes it difficult to separate retriever and generator performance in RAG evaluation.

In other domains, such as finance, VQAonBD \cite{raja_icdar_2023} focuses also only on tables, but FinQA \cite{chen_finqa_2021}, ConvFinQA \cite{chen_convfinqa_2022}, and TAT-DQA \cite{zhu_complex_2022} incorporate both text-and-tabular data from financial reports. Nonetheless, all financial datasets contain mainly context-dependent questions.

Moreover, several datasets are not publicly available, such as FinDER \cite{choi_finder_2025} and BioTABQA \cite{luo_biotabqa_2022}, or represent tables as images rather than structured text in markdown format \cite{tito_document_2021, pramanick_spiqa_2024}.
Other datasets are cross-domain, such as TableBench \cite{wu_tablebench_2025}, which provides multi-domain table QA for oracle context evaluation, while the UDA benchmark \cite{hui_uda_2024} aggregates multiple datasets. However, both remain limited by context-dependent questions. 
T$^2$-RAGBench closes this gap by providing a benchmark that focuses on text-and table-data, has no data contamination, and contains only context-independent questions.

\paragraph{RAG on Text-and-Table.}
RAG shows promise on text \cite{lewis_retrievalaugmented_2020}, but text-and-table evaluation is limited. THoRR \cite{kim_thorr_2024} simplifies tables via header-based retrieval, complementing ERATTA \cite{roychowdhury_eratta_2024}, which uses modular prompts and SQL for enterprise data. FinTextQA \cite{chen_fintextqa_2024} evaluates full RAG pipelines.
FinTMMBench \cite{10.1145/3746027.3755723} adds multi-modal and temporal RAG via dense/graph retrieval. Robust RAG \cite{joshi_robust_2024} links text, tables, visuals via image-based VLLMs, though less flexible than text methods. Despite progress, most works \cite{asai_selfrag_2024, gao_precise_2023, gao_retrievalaugmented_2023} test only a few RAG baselines, limiting generalizability.
\section{Task Definition}\label{task-def}
To clarify the task addressed by our benchmark, we define the following problem to be solved.

\begin{table*}[!t]
\centering
\resizebox{\textwidth}{!}{
\begin{tabular}{lrrrrrrrrr}
\toprule
\multirow{2}{*}{\textbf{Subset}} & 
\multirow{2}{*}{\textbf{Domain}} & 
\multirow{2}{*}{\textbf{PDF Source}} & 
\multicolumn{3}{c}{\textbf{\#Documents}} & 
\multicolumn{2}{c}{\textbf{\#QA Pairs}} &
\multicolumn{2}{c}{\textbf{Avg. Question Tokens}} \\
\cmidrule(lr){4-6} \cmidrule(lr){7-8} \cmidrule(lr){9-10}
& & & Original & Extracted & Avg. Token & Original & Generated & Original & Generated \\
\midrule
FinQA      & Finance & FinTabNet & 2,789 & 2,789  & 950.4 & 8,281  & 8,281  & 21.1  & 39.2 \\
ConvFinQA  & Finance & FinTabNet & 2,066 & 1,806  & 890.9 & 14,115 & 3,458  & 17.8  & 30.9 \\
TAT-DQA    & Finance & TAT-DQA   & 2,758 & 2,723  & 915.3 & 16,558 & 11,349 & 17.8  & 31.7 \\
\midrule

\textbf{Total} & \textbf{Finance} & \textbf{Multiple} & 7,613 & \textbf{7,318} & 924.2 & 38,954 & \textbf{23,088} & 19.0 & 34.3 \\
\bottomrule
\end{tabular}
}
\caption{
Comparison of original and generated QA pairs, documents, and average question and context lengths across T$^2$-RAGBench subsets. FinQA \cite{chen_finqa_2021} and ConvFinQA \cite{chen_convfinqa_2022} use FinTabNet \cite{9423317} as their PDF source, while TAT-DQA \cite{zhu_complex_2022} uses its own dataset. Avg. token count based on Llama 3.3 tokenizer.
}
\label{tab:merged_dataset_stats}
\end{table*}

\paragraph{Problem Formulation.}
The benchmark evaluates both the retrieval function \(f\) and the reasoning model \(M\) to optimize answer accuracy and efficiency in the unknown-context text-and-table QA setting.
We denote the user’s question by \(Q\) and the corresponding ground truth answer by \(A\). The evidence comes from two modalities: a segment of text content and a structured table, which we consider together as a single context entity denoted by \(C\). Thus, our entire context corpus is defined as \(\mathcal{C} = \{C_i\}\).
The task is divided into two stages:
\\
\textbf{Retrieval:}
A function
\begin{equation}
   f: \mathcal{C} \times Q \mapsto [C^*_k]_{k=1}^n
\end{equation}
selects the top-\(n\) most relevant context entities from the corpus \(\mathcal{C}\) for a given question \(Q\).
\\
\textbf{Answer Extraction:}
A language model
\begin{equation}
    M: \bigl([C^*_k]_{k=1}^n,\, Q\bigr) \mapsto A^*
\end{equation}
generates an answer \(A^*\) by reasoning over the retrieved text and tables.
\\
\textbf{Number Match:}
Numerical reasoning is evaluated using a new metric.
It allows for minor deviations and unit scale shifts. Let \(A^*\) and \(A\) be the predicted and ground truth answers, and denote their absolute values as \(a^* = |A^*|\) and \(a = |A|\).
\\
Given a tolerance threshold \(\varepsilon > 0\), the prediction is considered correct if either \(a^* < \varepsilon\) and \(a < \varepsilon\), or
\(|q - 1| < \varepsilon \) where
\[
q = \frac{a^*}{a} \cdot 10^{-\text{round}(\log_{10}(a^* / a))}.
\]
Here, \(\text{round}\) denotes rounding to the nearest integer.
This metric ensures robustness to rounding errors and magnitude scaling.

\paragraph{Retrieval Metrics.}
Let 
\[
\mathcal{D} = \{(Q_i, A_i, C_i)\}_{i=1}^N
\]
represent our dataset, where each tuple \((Q_i, A_i, C_i)\) consists of a question \(Q_i\), its unique ground-truth answer \(A_i\), and the corresponding unique ground-truth context \(C_i\).
Define the retrieval output:
\begin{equation}
R_i = f(\mathcal{C}, Q_i) = [C_{i,1}^*, C_{i,2}^*, \dots, C_{i,n}^*].
\end{equation}
The true rank is given by
\begin{equation}
r_i = \min\{k \mid C^*_{i,k} = C_i\}.
\end{equation}
We consider the Mean Reciprocal Rank at \(k\) (MRR@k), which focuses on the relevance of the top \(k\) retrieved contexts. It is defined as
\begin{equation}
\mathrm{MRR@}k = \frac{1}{N}\sum_{i=1}^N \frac{1}{r_i} \cdot \mathbb{I}(r_i \leq k),
\end{equation}
where \(\mathbb{I}(\cdot)\) is the indicator function, valued at 1 if the condition is met (i.e., \(r_i \leq k\)), and 0 otherwise.

\section{\texorpdfstring{T\textsuperscript{2}-RAGBench}{T2-RAGBench}}\label{sec:benchmark}

To construct our benchmark for text-table data suitable for RAG evaluation, we first surveyed existing datasets, as summarized in Table~\ref{tab:dataset_comparison}. 
As none fully met our criteria, we selected FinQA~\cite{chen_finqa_2021}, ConvFinQA~\cite{chen_convfinqa_2022}, and TAT-DQA~\cite{zhu_complex_2022} and restructured them to context-independent questions.


A question is considered context-independent if it has exactly one correct answer, even without access to \(\mathcal{C}\).
For all selected datasets, we applied custom preprocessing steps and reformulated questions using Llama 3.3-70B\footnote{\label{llama-footnote}\href{https://huggingface.co/kosbu/Llama-3.3-70B-Instruct-AWQ}{huggingface.co/kosbu/Llama-3.3-70B-Instruct-AWQ}} to ensure context-independence.

Each benchmark sample is a triple \((Q, A, C)\), where \(Q\) is a question, \(A\) the answer, and \(C\) the context composed of both text and table. Since all triples originate from \textit{ Oracle Context} settings, we assume that all required information to answer \(Q\) is fully contained within \(C\), and only within \(C\).
Table~\ref{tab:merged_dataset_stats} provides a detailed breakdown of the three subsets of T$^2$-RAGBench. While FinQA and ConvFinQA are based on FinTabNet, TAT-DQA is based on its own financial documents. The subsets consist of 1,806 to 2,789 documents, with each containing between 3,458 and 11,349 QA pairs. 
We included samples for each subset in Appendix~\ref{apx:dataset_samples}.

\subsection{Data Preparation}\label{preparation}
All subsets required tailored preprocessing to align with the requirements of our benchmark. FinQA is a numerical QA dataset based on financial reports from FinTabNet. We used it with company metadata and standardized all answer formats. ConvFinQA extends FinQA by adding multi-turn questions. We filtered only to include first-turn questions and normalized the answers for consistency. 
TAT-DQA is an independent dataset with diverse answer types. We filtered it to keep only numerical questions and normalized answer formats. Full details can be found in Appendix~\ref{apx:data-prep}.

\subsection{Data Creation}\label{filtering}

Following the preprocessing, the context-independent questions were generated.
First, the questions were reformulated using an
LLM. Subsequently, both quantitative and qualitative analyses were performed to verify that (1) the data quality remained consistent with the original, and (2) the reformulation process produced genuinely context-independent questions.

\paragraph{Question Reformulation.}


To generate context-independent questions, the original questions were reformulated, but the answers remained unchanged to preserve human-annotated quality.
For each of the 23,088 samples, a new question was generated using Llama 3.3-70B\footref{llama-footnote} with temperature = 0.7. The generation process was conducted by incorporating meta-information, such as company name, sector, and report year, which were not included in the original document.
The exact prompting template is detailed in Appendix~\ref{apx:reformat_prompt}.

\paragraph{Quantitative Analysis.}

\begin{figure}[!t]
    \centering
    \includegraphics[width=1\columnwidth]{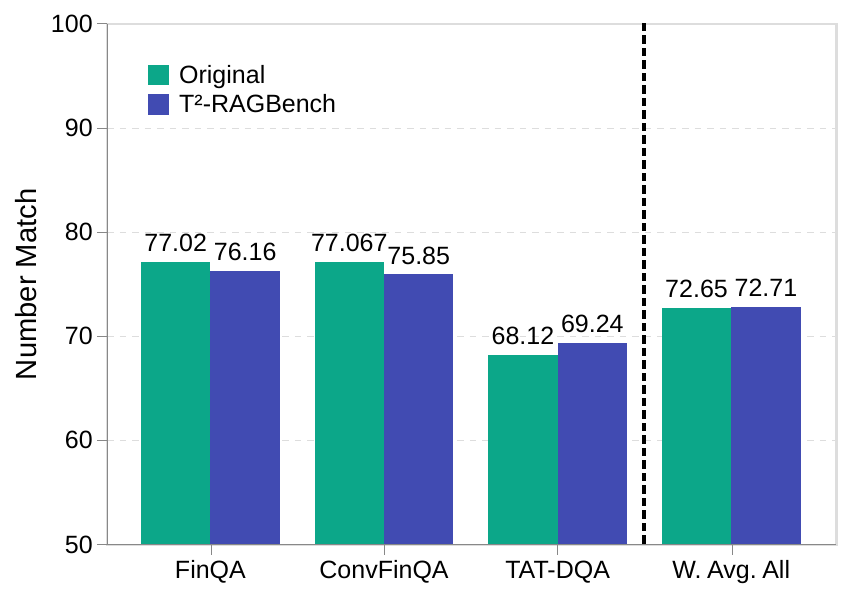}
    \caption[data statistics]{Number Match comparison per subset and weighted average all between original and reformulated questions from our new benchmark.}
    \label{fig:Statistics}
    \vspace{-1em}
\end{figure}

To verify that the rephrased questions remain consistent with the original answer,
we conducted a quantitative comparison of the original and reformulated questions across all subsets using Llama 3.3-70B\footref{llama-footnote} and \textit{Oracle Context}, as presented in Figure~\ref{fig:Statistics}.


Since the context is given, only Number Match was used to evaluate the QA pairs.
The accuracy between original and generated questions shows minimal deviation, with maximal differences of
2\% per subset and in average $<0.05\%$.
The ability of the LLM to answer the reformulated questions indicates that they retain the essential information required for numerical reasoning.

\paragraph{Human Validation.}

\begin{figure}[!t]
    \centering
    \includegraphics[width=1\columnwidth]{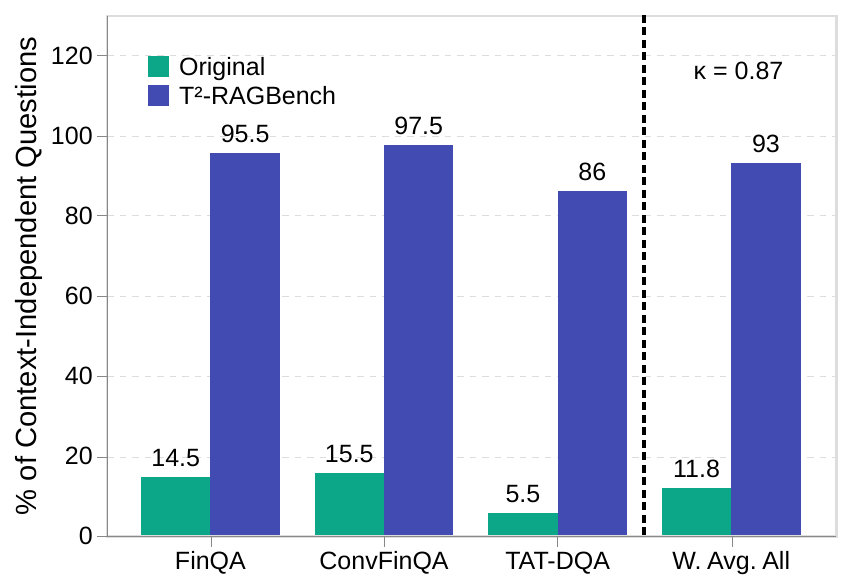}
    \caption[data statistics]{Percentage of context-independent questions (100 per subset, weighted avg overall). $\kappa$ indicates inter-annotator agreement.}
    \label{fig:human_agreement}
    \vspace{-1em}
\end{figure}

To further investigate whether the questions are now context-independent after reformulation, we conducted a human evaluation after the quantitative analysis.
Therefore, a random sample of 100 original and generated QA pairs per subset was manually labeled via a custom annotation tool (Appendix~\ref{apx:annotation_tool}). Each of the four financial experts annotated 200 samples from two different subsets, assessing whether the original questions were context-independent or context-dependent. 
The analysis reveals that only 11.8\% of questions in the original dataset were context-independent, compared to 93\% in the reformulated version (see Figure~\ref{fig:human_agreement}). 
This ensures that nearly all of the newly created QCA triples are suitable for RAG evaluation. 
Cohen’s Kappa was calculated to assess inter-annotator agreement, yielding an overall value of 0.87, indicating almost perfect agreement. Notably, only $1/3$ of the uncertain cases involved reformulated questions, suggesting that most ambiguity stemmed from original question formulations. For better transparency, we include representative disagreement examples in Appendix~\ref{apx:annotation_samples} and in our repository\footref{repo}.


\subsection{Data Statistics}\label{statistic}
Table~\ref{tab:merged_dataset_stats} presents an overview of the dataset. It comprises 7,318 real-world documents with an average length of 924.2 tokens.
In total, T$^2$-RAGBench consists of 23,088 QCA triples extracted from roughly 40k questions.
Questions increased by \textasciitilde15 tokens with added semantic details (e.g., company names, years), making them context-independent and suitable for RAG evaluation. All other parameters (Metadata, IDs, etc.) of the dataset remained the same.


%
\section{Experiments}
\label{sec:results}


To evaluate the suitability of our benchmark for RAG methods, we report results across all subsets using various models and RAG approaches. 
This section describes the setup (Section~\ref{setup}), compares methods (Section~\ref{methods}), defines metrics (Section~\ref{evaluation-metric}), and presents the main results (Section~\ref{sub:main-results}), revealing a large gap between oracle and current RAG performance. We then analyze this gap via two ablations (Section~\ref{sub:ablation}) and a manual error analysis (Section~\ref{sub:error}).

\subsection{Experimental Setup}\label{setup}
For the evaluation of the benchmark, each subset was evaluated independently. 
First, all contexts were \review{transformed} into markdown format and uniquely stored into a Chroma~\cite{chromateam_chroma_2025} vector db using the embeddings created with the multilingual e5-large instruct model~\cite{wang_multilingual_2024}, having an embedding size of 1024.
That was done for all RAG methods except for Summarization and \review{SumContext}, where the summarized context was embedded.
A retrieval query was used to retrieve from the \review{embedding} model (See Appendix~\ref{apx:retrieval_prompt}).
The Top-3 documents were selected and passed to the generator in the main evaluation.
As generators, we employed quantized LLaMA 3.3 70B\footref{llama-footnote}, a decoder-only transformer, and QwQ-32B\footnote{\label{qwq}\href{https://huggingface.co/Qwen/QwQ-32B-AWQ}{hugginface.co/Qwen/QwQ-32B-AWQ}}, to evaluate performance across 
multiple model architectures on two NVIDIA H100.
Due to resource limitations, we utilize quantized models, which exhibit negligible performance loss~\cite{jin_comprehensive_2024}.
The prompt template is provided in Appendix~\ref{epx:evaluation_prompt}.

\subsection{RAG Methods}\label{methods}
The following section briefly describes all evaluated RAG methods to show the SOTA performance on T$^2$-RAGBench, categorized by the retrieval complexity and augmentation strategy.

\paragraph{Pretrained-Only and Oracle Context.}
In the \emph{Pretrained-Only} setup, no retriever is employed, and models must answer questions solely based on their pretraining knowledge. Conversely, the \emph{Oracle Context} setting assumes that the relevant context is directly \review{passed} to the generator.

\paragraph{Basic RAG Methods.}
This category includes approaches that retrieve documents using standard embedding-based methods. The \emph{Base RAG} implementation follows the original RAG approach~\cite{lewis_retrievalaugmented_2020}, where only the question is embedded to retrieve the top-k documents, which are then passed unchanged to the generator. \emph{Hybrid BM25}~\cite{gao_complement_2021} combines sparse lexical retrieval using BM25 with dense vector retrieval, leveraging both methods to improve recall and relevance. Additionally, the \emph{Reranker} method~\cite{glass-etal-2022-re2g} applies a cross-encoder model\footnote{\href{https://huggingface.co/cross-encoder/ms-marco-MiniLM-L6-v2}{huggingface.co/cross-encoder/ms-marco-MiniLM-L6-v2}} after initial retrieval to reorder documents based on their relevance in a shared embedding space.

\paragraph{Advanced RAG Methods.}
This category consists of methods that modify the query, transform retrieved contexts, or employ iterative retrieval strategies. The \emph{HyDE} method~\cite{gao_precise_2023} generates hypothetical answers for each question, using them as refined queries to retrieve more relevant documents (For prompt see Appendix~\ref{apx:hyde_prompt}). \emph{Summarization} reduces noise by summarizing each retrieved context using an LLM, focusing on essential information. \emph{SumContext} applies the similar summarization step but retains the original full documents for generation, aiming to reduce distractions while preserving content fidelity (See Appendix~\ref{apx:summarize_prompt}).

\begin{table*}[!t]
\centering
\resizebox{\textwidth}{!}{
\begin{tabular}{>{\raggedright}p{2.5cm}l|ccc|ccc|ccc|ccc}
\toprule
\multirow{2}{*}{\textbf{Model}} & \multirow{2}{*}{\textbf{RAG Method}} 
  & \multicolumn{3}{c}{\textbf{FinQA}}  
  & \multicolumn{3}{c}{\textbf{ConvFinQA}}  
  & \multicolumn{3}{c}{\textbf{TAT-DQA}}  
  & \multicolumn{3}{c}{\textbf{W. Avg Total}} \\
\cmidrule(lr){3-5} 
\cmidrule(lr){6-8} 
\cmidrule(lr){9-11} 
\cmidrule(lr){12-14} 
& & NM & MRR@3 & \review{R@3} & NM & MRR@3 & \review{R@3} & NM & MRR@3 & \review{R@3} & \review{NM} & \review{MRR@3} & \review{R@3} \\
\midrule
\multirow{7}{*}{\centering\makecell{Llama 3.3-70B\\ + Multilingual\\E5-Large\\Instruct }} 
   & \textit{+ Pretrained-Only}     & 7.9 & --    & \review{--}   & 2.8 & --     & \review{--}   & 3.7 & --    & \review{--}   & \review{5.1} & \review{--} & \review{--} \\
   & \textit{+ Oracle Context}      & \review{76.2} & 100 & \review{100} & 75.8 & 100  & \review{100} & 69.2 & 100 & \review{100} & \review{72.7} & \review{100} & \review{--} \\
   \cmidrule(lr){2-14}
   & \textit{+ Base‐RAG}       & 39.5 & 38.7 & \review{49.7} & 47.4 & 42.2 & \review{53.8} & 29.6 & 25.2 & \review{28.4} & \review{35.8} & \review{32.6} & \review{39.8} \\
   & \textit{+ Hybrid BM25}    & 41.7 & 40.0 & \review{53.0} & 50.3 & 43.5 & \review{57.2} & \textbf{37.4} & \textbf{29.2} & \review{\textbf{44.4}} & \review{\textbf{40.9}} & \review{35.2} & \review{\textbf{49.4}} \\
   & \textit{+ Reranker}       & 32.4 & 29.0 & \review{36.2} & 37.3 & 32.3 & \review{40.5} & 27.0 & 22.8 & \review{28.4} & \review{30.5} & \review{26.4} & \review{33.0} \\
   \cmidrule(lr){2-14}
   & \textit{+ HyDE}           & 38.4 & 35.4 & \review{45.7} & 44.8 & 39.8 & \review{50.9} & 26.7 & 20.8 & \review{26.7} & \review{33.6} & \review{28.9} & \review{37.1} \\
   & \textit{+ Summarization}  & 27.3 & 47.3 & \review{\textbf{59.5}} & 35.2 & 52.1 & \review{\textbf{63.8}} & 14.6 & 24.7 & \review{31.5} & \review{22.2} & \review{\textbf{36.9}} & \review{\underline{46.4}} \\
   & \textit{+ SumContext}     & \textbf{47.2} & \textbf{47.3} & \review{\textbf{59.4}} & \textbf{55.5} & \textbf{52.1} & \review{\textbf{63.8}} & 29.1 & 24.8 & \review{31.4} & \review{\underline{39.5}} & \review{\textbf{37.0}} & \review{\underline{46.3}} \\
\midrule
\midrule
\multirow{7}{*}{\centering\makecell{QwQ-32B\\ + Multilingual\\E5-Large\\Instruct }} 
   & \textit{+ Pretrained-Only}     & 7.5 & --    & \review{--}   & 2.4 & --   & \review{--}   & 4.4 & --    & \review{--}   & \review{5.2} & \review{--} & \review{--} \\
   & \textit{+ Oracle Context}      & 72.4 & 100 & \review{--} & 85.4 & 100  & \review{--} & 71.1 & 100 & \review{--} & \review{73.7} & \review{100} & \review{--} \\
   \cmidrule(lr){2-14}
   & \textit{+ Base-RAG}            & 39.6 & 38.7 & \review{49.7} & 48.7 & 42.4 & \review{53.8} & 27.9 & 25.2 & \review{28.4} & \review{35.2} & \review{32.6} & \review{39.8} \\
   & \textit{+ Hybrid BM25}         & 41.8 & 39.8 & \review{53.0} & 51.6 & 43.6 & \review{57.2} & \textbf{37.2} & \textbf{29.3} & \review{\textbf{44.4}} & \review{\textbf{41.0}} & \review{35.2} & \review{\textbf{49.4}} \\
   & \textit{+ Reranker}            & 30.8 & 29.0 & \review{36.2} & 37.5 & 32.7 & \review{40.5} & 25.6 & 22.9 & \review{28.4} & \review{29.2} & \review{26.6} & \review{33.0} \\
   \cmidrule(lr){2-14}
   & \textit{+ HyDE}                & 36.8 & 35.4 & \review{45.7} & 45.7 & 39.9 & \review{50.9} & 24.7 & 20.7 & \review{26.7} & \review{32.2} & \review{28.8} & \review{37.1} \\
   & \textit{+ Summarization}       & 26.9 & 47.2 & \review{\textbf{59.5}} & 35.6 & 52.2 & \review{\textbf{63.8}} & 13.9 & 24.7 & \review{31.5} & \review{21.8} & \review{\textbf{36.9}} & \review{\underline{46.4}} \\
   & \textit{+ SumContext}          & \textbf{45.6} & \textbf{47.3} & \review{\textbf{59.4}} & \textbf{56.9} & \textbf{52.2} & \review{\textbf{63.8}} & 27.3 & 24.7 & \review{31.4} & \review{\underline{38.3}} & \review{\textbf{36.9}} & \review{\underline{46.3}} \\
\bottomrule
\end{tabular}
}
\caption{\review{Overall performance (Number Match (NM), MRR@3, and R@3) of both models on T$^2$-RAGBench. Number Match represents the percentage of correctly answered questions based on their numerical representation. R@3 and MRR@3 evaluate retrieval effectiveness. Cells in \textbf{Bold} indicate the highest value over all RAG methods, and \underline{underlined} indicate the best value across RAG method categories.}}
\label{tab:main_results}
\end{table*}

\subsection{Evaluation Metrics}\label{evaluation-metric}

We use Number Match and MRR@$k$ as our main metrics as defined in Section~\ref{task-def}, but also report Recall@$k$ for better comparability and transparency.
\textbf{Number Match} evaluates if a numerical prediction closely matches the gold numerical answer. It compares predicted and ground truth values using relative tolerance ($\epsilon = 1\mathrm{e}{-2}$), accounting for scale invariance. Non-numeric predictions or mismatches are considered incorrect.
For \textbf{MRR} and \textbf{Recall}, we choose $k=3$, \review{which} measures whether the first relevant document appears in the top-3 retrieved results, rewarding higher ranks for MRR.
\review{We limit evaluation to 3 documents, as the average length is \review{924.2} tokens. Increasing the number of documents increases input size, slows inference, and hinders LLM performance, making it impractical \cite{li_long_2024}.
}

\subsection{\review{Main Results}}\label{sub:main-results}
\review{
This section discusses our main results presented in Table~\ref{tab:main_results} for all three evaluation subcategories.
}
\paragraph{Pretrained-Only and Oracle Context.}
The results from the \textit{Pretrained-Only} setting show that across all subsets, the questions cannot be answered directly from the models’ pretraining data. This highlights the importance of RAG and the need for a dedicated benchmark. While reformulated questions may resemble seen content, especially since most S\&P 500 reports predate 2023, this applies to both foundation and reasoning models.
In contrast, the \textit{Oracle Context} setting shows consistently high performance on Number Match across all subsets and both models, highlighting both the strong numerical reasoning abilities of the models and the feasibility of the task for modern LLMs in this setting. Notably, there is no significant performance difference between Llama and QwQ ($<0.3\%$).

\paragraph{Base RAG Methods.}
\review{
For base RAG methods, the benchmark shows that all SOTA models still struggle to match the performance achieved in \textit{Oracle Context}.
}
Nevertheless, this benchmark offers the possibility to compare the different methods \review{precisely}. 
For \textit{Base-RAG}, MRR@3 \review{and R@3} averaging below 40\%, meaning relevant documents are often missing in the top-3, which leads to a significant drop in Number Match.
This effect is particularly evident in TAT-DQA, where, despite having a similar number of documents as FinQA, relevant information is harder to retrieve for all tested methods. 
\textit{Hybrid BM25} consistently outperforms base RAG in \review{Number Match, MRR@3, and R@3} on average. Interestingly, the Reranker performs worse than \textit{Base} and \textit{Hybrid BM25} RAG methods, suggesting that the reranking model \review{struggles with} text-and-table data.

\paragraph{Advanced RAG Methods.}
One way to improve the performance of RAG methods is to improve the linking of the query with the context. 
However, \textit{HyDE} shows even a drop in performance in MRR@3 \review{and R@3} across all subsets in comparison to the \textit{Base-RAG}. This may be due to the models’ difficulty in generating well-structured content matching the format of the documents, which often include both text and tables.


The \textit{Summarization} approach performed well on MRR@3 for FinQA and ConvFinQA by condensing relevant information and removing noise. However, it underperforms on TAT-DQA, warranting further investigation. In general, this often led to a drop in NM, as essential information needed to answer the questions was also lost during summarization.
\textit{SumContext} retrieves from a summarized context but provides the full original context. This approach improved MRR@3 while maintaining stable NM, achieving an average NM of 37.4\% resp. 36.7\%. Nevertheless, the performance does not improve across all subsets, indicating strong sensitivity to prompts and datasets.
\review{
Interestingly, MRR@3 is 1.8\% higher than \textit{Hybrid BM25}, despite lower R@3, suggesting retrieved documents are ranked higher in Summarization and SumContext.
}

\subsection{Ablation Studies}\label{sub:ablation}

\begin{table}[!t]
\centering
\resizebox{\columnwidth}{!}{%
\begin{tabular}{l ccc}
\toprule
\textbf{\review{Embedding Model}} & \textbf{\review{R@1}} & \textbf{\review{R@5}} & \textbf{\review{MRR@5}} \\
\midrule
\review{Stella-EN-1.5B} & \review{2.2} & \review{5.2} & \review{3.3} \\
\review{GTE-Qwen2 1.5B Instruct} & \review{12.5} & \review{27.6} & \review{18.0} \\
\review{Multilingual E5-Instruct} & \textbf{\review{26.4}} & \textbf{\review{49.7}} & \textbf{\review{35.1}} \\
\midrule
\review{Gemini: Text-Embedding-004} & \review{32.3} & \review{53.6} & \review{41.7} \\
\review{OpenAI: Text-Embedding-3 Large} & \textbf{\review{34.6}} & \textbf{\review{57.4}} & \textbf{\review{44.7}} \\
\bottomrule
\end{tabular}
}
\caption{Retrieval performance of embedding models on T\textsuperscript{2}-RAGBench using \textit{Base-RAG} with $k=5$ retrieved documents, evaluated on R@1, R@5 and MRR@5. Scores are weighted avg. over all subsets. Model descriptions are in Appendix~\ref{apx:retrieval_models}.}
\label{tab:retrieval_models}
\vspace{-1.5em}
\end{table}

\paragraph{Embedding Models.}
We evaluate various embedding models with the \textit{Base-RAG} approach to assess their impact on retrieval performance. As shown in Table~\ref{tab:retrieval_models}, among the open-source models, \textit{Multilingual E5-Instruct} performs best, achieving 29.4\% R@1 and 38.6 MRR@5. The closed-source models perform slightly better, with \review{the} OpenAI model reaching the highest R@1 of 33.8\% and MRR@5 of 43.6. However, none of the models, regardless of model size, achieve satisfactory performance on the challenging text-and-table setting at R@1, indicating that retrieving the correct document remains a core challenge in T\textsuperscript{2}-RAGBench, \review{because text-and-table documents seem to be challenging for SOTA embedding models.}

\begin{figure}[!t]
    \centering
    \includegraphics[width=.9\columnwidth]{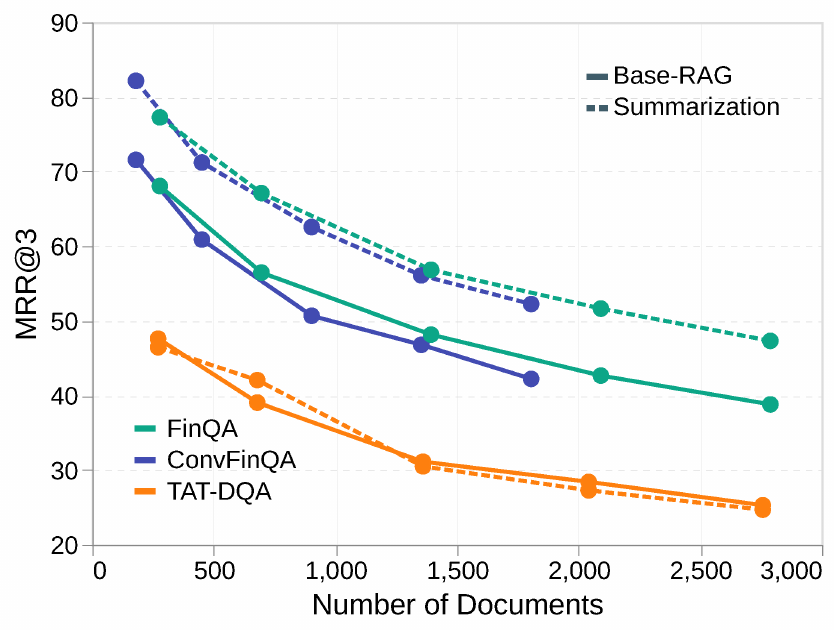}
    \caption[data statistics]{\review{MRR@3 comparison for FinQA, ConvFinQA, and TAT-DQA across five evenly split document subsets.}}
    \label{fig:number_documents}
\end{figure}

\paragraph{Number of Documents.}  
Figure~\ref{fig:number_documents} shows how retrieval performance changes with the number of documents for \textit{Base-RAG} and \textit{Summarization}, using 5 random percentage ascending subsets per dataset. Two main findings emerge: (1) MRR@3 drops below 50\% with 3K documents, meaning the correct document appears in the top 3 only half the time; (2) Summarization improves results for FinQA and ConvFinQA, performs similarly on TAT-DQA, where summarizing tabular content is more challenging.

\subsection{Manual Error Analysis}\label{sub:error}
\begin{figure}[!t]
    \centering
    \includegraphics[width=0.9\columnwidth]{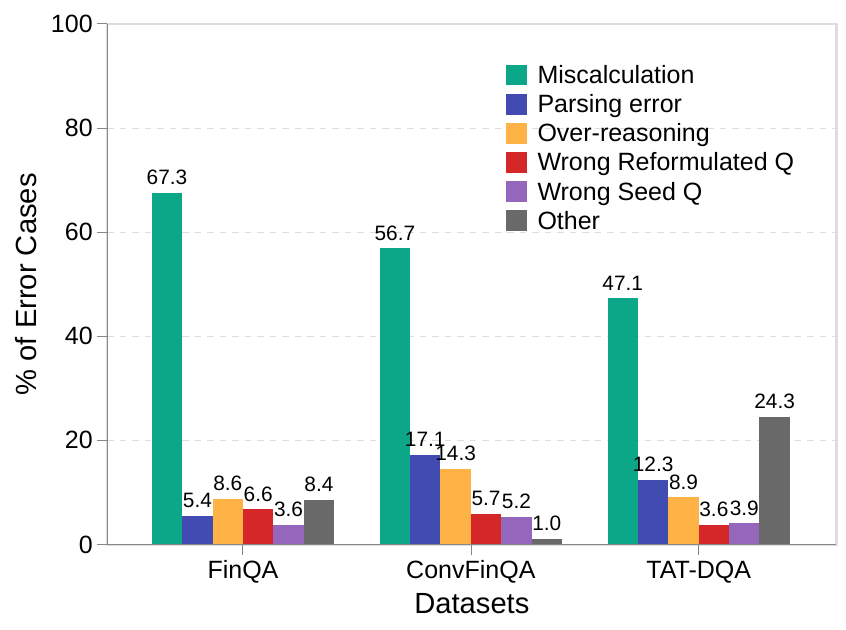}
    \caption[data statistics]{\review{Results of the manual error analysis. Percentage of each error category per subset.}}
    \label{fig:error}
\end{figure}

\review{
We performed a manual qualitative error analysis on 25\% of the \textit{Oracle Context} errors from our main results, comprising 1,583 annotated cases across all subsets (see Figure~\ref{fig:error}). Each error was categorized into one of six categories: miscalculation, parsing error, over-reasoning, wrong reformulated question, wrong seed question, and other (see Appendix~\ref{apx:error_categories} for more information). 
}

\review{
The majority of error cases arise from arithmetic mistakes, parsing errors, or instances of unnecessary reasoning, indicating that models continue to struggle with reliably answering certain types of questions. A common failure involves inserting incorrect values into tables or producing arithmetic results that deviate slightly from the correct answer. This pattern is consistent across all three subsets, suggesting that such challenges persist irrespective of the underlying data source.
}
\review{
Additionally, approximately 6\% of errors in each subset are attributed to reformulation failures. In nearly 90\% of these, the metric changed from 'value' to 'percentage value,' which confuses the generator. Approximately 5\% of errors originate from unclear or ambiguous seed questions. Other errors include parsing issues and outputs with only NA values (especially for TAT-DQA), making diagnosis difficult. Overall, the benchmark remains challenging, with room for improvement in generation, but most questions remain suitable for evaluating RAG.
}

\subsection{Main Takeaways}
Overall, our results show that even the strongest RAG method examined (\textit{Hybrid BM25}) falls short of \textit{Oracle Context} performance in NM by almost 30\%. 
This performance gap underscores the benchmark's ability to quantify retrieval effectiveness and highlights the remaining challenges in achieving oracle-level performance with RAG. 
Even when using other RAG methods like Hybrid BM25, the performance can only be improved by \review{2.5\%} on average on MRR and \review{5\%} in comparison to \textit{Base-RAG}. 
We further analyzed the impact of other factors and find that even SOTA retrieval models achieve less than 50\% MRR@5, highlighting that RAG on text-and-table data remains challenging; additionally, retrieval performance with 3K documents reveals that this task still offers significant room for improvement.

\section{Conclusion}
\label{sec:conclusion}

In this paper, we introduced our newly created benchmark, \textbf{T$^2$-RAGBench}, which contains \review{23,088} question-answer-context triples. It includes questions derived from over \review{7,318} documents and is designed to evaluate RAG methods for numerical reasoning over text-table data in the unknown context setting. 
While other datasets are defined in an \textit{Oracle Context}, our benchmark uses context-independent question making it possible to evaluate RAG methods. We demonstrate that our benchmark meets its intended goals through quantitative analysis and human validation.
We test multiple RAG methods on the benchmark and find that \textit{Hybrid BM25}, which combines dense and sparse retrieval, performs best. Additionally, we conducted ablation studies showing that current SOTA embedding models achieve low R@5 and MRR@5 scores on text-and table contexts.  
With T$^2$-RAGBench, we aim to facilitate the development of more RAG methods suitable for text-and-table documents, supporting the creation of real-world systems that can automatically analyze complex documents.
Given the dataset's size and complexity, we do not expect SOTA models to solve the benchmark in the near future. 
In future work, we plan to evaluate additional RAG methods to identify which factors have the greatest impact on text-and-table data. Adding more data from other domains is also necessary to further generalize the evaluation.
\section*{Limitations}
\label{sec:limitations}

This section outlines the key limitations related to the methodology and dataset that may affect the validity and generalizability of the presented results.

\paragraph{Lack of Human Verification and Authenticity.} 
The questions used in the benchmark were generated synthetically, which can lead to distortions, as models do not inevitably generate the type of questions that real-world users would ask. Therefore, transferability to real systems may be affected. 
\review{
Although humans annotated the original question-answer pairs, there is no definitive guarantee that the generated questions will be formulated in a way that allows other models to answer them equivalently.
}

Another point is that a comprehensive verification process was only partly conducted on the benchmark questions.
While we verified 100 samples per subset with four annotators in the benchmark, that the benchmark fulfills the requirements to be an evaluation dataset for our proposed task. Nevertheless, they can still be some questions that are not suitable to find the right context.

\paragraph{Domain-Specific Application.} 
The presented work aims to present a benchmark that can test text-table datasets from different document types with different knowledge. Nevertheless, the dataset consists only of financial documents that have 
the same standardized structure, consistent terminology, and domain-specific content. As a result, the model's performance is tailored to this domain and can only be partly assumed to generalize to other types of document layouts or content types, such as medical reports, scientific publications, or administrative forms, where table-text relationships can vary significantly. Still, given the wide-ranging application of financial reporting standards, our work contributes to this specific domain.

\paragraph{Use of Quantized Models.} 
\review{
Due to limited resources, all evaluations were conducted using quantized versions of the models, which enabled faster inference times and the execution of large open-source models. While quantization offers clear advantages in terms of computational efficiency, it often comes at the cost of reduced numerical precision and model accuracy. Therefore, the performance may be lower than that of full-precision SOTA models. However, since the focus of this paper is on comparing suitable RAG methods, we consider this negligible.
}

\section*{\review{Ethical Considerations}}
\review{
This work introduces a benchmark dataset constructed from publicly available financial documents. All data used originates from previously published datasets (FinQA, ConvFinQA, TAT-DQA), which are either publicly accessible or sourced from publicly available company reports. No private, confidential, or personally identifiable information is included. The reformulated questions were synthetically generated using LLMs and subsequently validated by experts to ensure quality and context-independence. Human evaluation was conducted with informed consent and anonymized input. We acknowledge that while synthetic reformulation enhances benchmarking utility, it may not fully capture the natural distribution of user queries.
}

\section*{Acknowledgement}
This work is supported by the Genial4KMU project, Universität Hamburg, funded by BMBF (grant no. 16IS24044B).

\bibliography{clean}

\newpage
\appendix
\onecolumn

\section{Dataset Samples}\label{apx:dataset_samples}
In the following, we give two examples for each dataset subset, including the original question, the reformulated question, and the corresponding context. Due to the limited page width, we had to wrap the text of the context.

\begin{tcolorbox}[
  colback=gray!20,  
  colframe=gray!20, 
  boxrule=0pt,      
  sharp corners    
]
\textbf{Dataset / ID:}\\
train\_finqa2516 \\ \\
\textbf{Question:}\\
  what is the growth rate in net revenue from 2010 to 2011? \\
  \\
\textbf{Reformulated:}\\
  What was the percentage change in Entergy's net revenue from 2010 to 2011, considering the impact of the mark-to-market tax settlement sharing, retail electric price adjustments, and other factors as outlined in the 2011 financial discussion and analysis?  \\ \\
  \textbf{Context:}\\
\begin{Verbatim}[fontsize=\small,frame=none, baselinestretch=1, breaklines=true]
entergy louisiana , llc and subsidiaries management 2019s financial discussion and
analysis plan to spin off the utility 2019s transmission business see the 201cplan to spin
off the utility 2019s transmission business 201d section of entergy corporation and
subsidiaries management 2019s financial discussion and analysis for a discussion of this
matter , including the planned retirement of debt and preferred securities .results of
operations net income 2011 compared to 2010 net income increased $ 242.5 million primarily
due to a settlement with the irs related to the mark-to-market income tax treatment of
power purchase contracts , which resulted in a $ 422 million income tax benefit .the net
income effect was partially offset by a $ 199 million regulatory charge , which reduced
net revenue , because a portion of the benefit will be shared with customers .see note 3
to the financial statements for additional discussion of the settlement and benefit
sharing .2010 compared to 2009 net income decreased slightly by $ 1.4 million primarily
due to higher other operation and maintenance expenses , a higher effective income tax
rate , and higher interest expense , almost entirely offset by higher net revenue .net
revenue 2011 compared to 2010 net revenue consists of operating revenues net of : 1 ) fuel
, fuel-related expenses , and gas purchased for resale , 2 ) purchased power expenses ,
and 3 ) other regulatory charges ( credits ) .following is an analysis of the change in
net revenue comparing 2011 to 2010 .amount ( in millions ) ._|    |
| amount ( in millions )
||---:|:--------------------------------------|:-------------------------||  0 | 2010 net
revenue                      | $ 1043.7                 ||  1 | mark-to-market tax
settlement sharing | -195.9 ( 195.9 )         ||  2 | retail electric price
| 32.5                     ||  3 | volume/weather                        | 11.6
||  4 | other                                 | -5.7 ( 5.7 )             ||  5 | 2011 net
revenue                      | $ 886.2                  |_the mark-to-market tax
settlement sharing variance results from a regulatory charge because a portion of the
benefits of a settlement with the irs related to the mark-to-market income tax treatment
of power purchase contracts will be shared with customers , slightly offset by the
amortization of a portion of that charge beginning in october 2011 .see notes 3 and 8 to
the financial statements for additional discussion of the settlement and benefit sharing
.the retail electric price variance is primarily due to a formula rate plan increase
effective may 2011 .see note 2 to the financial statements for discussion of the formula
rate plan increase. .
  \end{Verbatim}
\end{tcolorbox}

\begin{tcolorbox}[
  colback=gray!20,  
  colframe=gray!20, 
  boxrule=0pt,      
  sharp corners    
]
\textbf{Dataset / ID:}\\
train\_finqa518 \\ \\
\textbf{Question:}\\
  at december 312008 what was the total liabilities acquired for this plan in millions \\
  \\
\textbf{Reformulated:}\\
  As of December 31, 2008, what was the total amount of liabilities acquired by Republic Services for the BFI post-retirement healthcare plan, as disclosed in their 2008 consolidated financial statements? \\ \\
  \textbf{Context:}\\
\begin{Verbatim}[fontsize=\small,frame=none, baselinestretch=1, breaklines=true]
estimated future pension benefit payments for the next ten years under the plan ( in
millions ) are as follows : estimated future payments: ._|    | 2009              |   $
14.9 ||---:|:------------------|---------:||  0 | 2010              |     15.9 ||  1 |
2011              |     16.2 ||  2 | 2012              |     19.2 ||  3 | 2013
|     21.9 ||  4 | 2014 through 2018 |    142.2 |_bfi post retirement healthcare plan we
acquired obligations under the bfi post retirement healthcare plan as part of our
acquisition of allied .this plan provides continued medical coverage for certain former
employees following their retirement , including some employees subject to collective
bargaining agreements .eligibility for this plan is limited to certain of those employees
who had ten or more years of service and were age 55 or older as of december 31 , 1998 ,
and certain employees in california who were hired on or before december 31 , 2005 and who
retire on or after age 55 with at least thirty years of service .liabilities acquired for
this plan were $ 1.2 million and $ 1.3 million , respectively , at the acquisition date
and at december 31 , 2008 .multi-employer pension plans we contribute to 25 multi-employer
pension plans under collective bargaining agreements covering union- represented employees
.we acquired responsibility for contributions for a portion of these plans as part of our
acquisition of allied .approximately 22% ( 22 % ) of our total current employees are
participants in such multi- employer plans .these plans generally provide retirement
benefits to participants based on their service to contributing employers .we do not
administer these multi-employer plans .in general , these plans are managed by a board of
trustees with the unions appointing certain trustees and other contributing employers of
the plan appointing certain members .we generally are not represented on the board of
trustees .we do not have current plan financial information from the plans 2019
administrators , but based on the information available to us , it is possible that some
of the multi-employer plans to which we contribute may be underfunded .the pension
protection act , enacted in august 2006 , requires underfunded pension plans to improve
their funding ratios within prescribed intervals based on the level of their underfunding
.until the plan trustees develop the funding improvement plans or rehabilitation plans as
required by the pension protection act , we are unable to determine the amount of
assessments we may be subject to , if any .accordingly , we cannot determine at this time
the impact that the pension protection act may have on our consolidated financial position
, results of operations or cash flows .furthermore , under current law regarding multi-
employer benefit plans , a plan 2019s termination , our voluntary withdrawal , or the mass
withdrawal of all contributing employers from any under-funded , multi-employer pension
plan would require us to make payments to the plan for our proportionate share of the
multi- employer plan 2019s unfunded vested liabilities .it is possible that there may be a
mass withdrawal of employers contributing to these plans or plans may terminate in the
near future .we could have adjustments to our estimates for these matters in the near term
that could have a material effect on our consolidated financial condition , results of
operations or cash flows .our pension expense for multi-employer plans was $ 21.8 million
, $ 18.9 million and $ 17.3 million for the years ended december 31 , 2008 , 2007 and 2006
, respectively .republic services , inc .and subsidiaries notes to consolidated financial
statements %%transmsg*** transmitting job : p14076 pcn : 133000000 ***%%pcmsg|131
|00027|yes|no|02/28/2009 21:12|0|0|page is valid , no graphics -- color : d| .
  \end{Verbatim}
\end{tcolorbox}

\begin{tcolorbox}[
  colback=gray!20,  
  colframe=gray!20, 
  boxrule=0pt,      
  sharp corners    
]
\textbf{Dataset / ID:}\\
TatQA 8e642bdce983286cbaffa9661d24157a \\ \\
\textbf{Question:}\\
  What was the total intrinsic value of RSUs which vested during 2019? \\
  \\
\textbf{Reformulated:}\\
  What was the total intrinsic value of RSUs that vested during the year ended March 31, 2019, for Microchip Technology Inc.? \\ \\
  \textbf{Context:}\\
\begin{Verbatim}[fontsize=\small,frame=none, baselinestretch=1, breaklines=true]
Microsemi Acquisition-related Equity AwardsIn connection with its acquisition of Microsemi
on May 29, 2018, the Company assumed certain restricted stock units (RSUs), stock
appreciation rights (SARs), and stock options granted by Microsemi. The assumed awards
were measured at the acquisition date based on the estimated fair value, which was a total
of $175.4 million. A portion of that fair value, $53.9 million, which represented the pre-
acquisition vested service provided by employees to Microsemi, was included in the total
consideration transferred as part of the acquisition. As of the acquisition date, the
remaining portion of the fair value of those awards was $121.5 million, representing post-
acquisition share-based compensation expense that will be recognized as these employees
provide service over the remaining vesting periods. During the year ended March 31, 2019,
the Company recognized $65.2 million of share-based compensation expense in connection
with the acquisition of Microsemi, of which $3.5 million was capitalized into inventory
and $17.2 million was due to the accelerated vesting of outstanding equity awards upon
termination of certain Microsemi employees.Atmel Acquisition-related Equity AwardsIn
connection with its acquisition of Atmel on April 4, 2016, the Company assumed certain
RSUs granted by Atmel. The assumed awards were measured at the acquisition date based on
the estimated fair value, which was a total of $95.9 million. A portion of that fair
value, $7.5 million, which represented the pre-acquisition vested service provided by
employees to Atmel, was included in the total consideration transferred as part of the
acquisition. As of the acquisition date, the remaining portion of the fair value of those
awards was $88.4 million, representing post-acquisition share-based compensation expense
that will be recognized as these employees provide service over the remaining vesting
periods.Combined Incentive Plan InformationRSU share activity under the 2004 Plan is set
forth below:|                             | Number of Shares | Weighted Average Grant Date
Fair Value ||-----------------------------|------------------|----------------------------
-----------|| Nonvested shares at March 31, 2016 | 6,307,742        | $36.76
|| Granted                     | 1,635,655        | 51.46
|| Assumed upon acquisition    | 2,059,524        | 46.57
|| Forfeited                   | (722,212)        | 43.58
|| Vested                      | (2,861,253)      | 38.60
|| Nonvested shares at March 31, 2017 | 6,419,456        | 42.06
|| Granted                     | 1,267,536        | 77.26
|| Forfeited                   | (279,051)        | 49.65
|| Vested                      | (1,735,501)      | 38.00
|| Nonvested shares at March 31, 2018 | 5,672,440        | 50.79
|| Granted                     | 1,951,408        | 77.83
|| Assumed upon acquisition    | 1,805,680        | 91.70
|| Forfeited                   | (408,242)        | 73.36
|| Vested                      | (2,729,324)      | 61.51
|| Nonvested shares at March 31, 2019 | 6,291,962        | $64.81
|The total intrinsic value of RSUs which vested during the years ended March 31, 2019,
2018 and 2017 was $229.3 million, $146.0 million and $166.1 million, respectively. The
aggregate intrinsic value of RSUs outstanding at March 31, 2019 was $522.0 million,
calculated based on the closing price of the Company’s common stock of $82.96 per share on
March 29, 2019. At March 31, 2019, the weighted average remaining expense recognition
period was 1.91 years.
  \end{Verbatim}
\end{tcolorbox}

\begin{tcolorbox}[
  colback=gray!20,  
  colframe=gray!20, 
  boxrule=0pt,      
  sharp corners    
]
\textbf{Dataset / ID:}\\
TatQA a210c0538af4df5f8881dcb8f1bf00ff \\ \\
\textbf{Question:}\\
  What was the Accrued compensation and employee benefits in 2018? \\
  \\
\textbf{Reformulated:}\\
  What was the accrued compensation and employee benefits for Jabil Circuit Inc. as of August 31, 2018? \\ \\
  \textbf{Context:}\\
\begin{Verbatim}[fontsize=\small,frame=none, baselinestretch=1, breaklines=true]
Intangible asset amortization for fiscal years 2019, 2018 and 2017 was approximately $31.9
million, $38.5 million and $35.5 million, respectively. The estimated future amortization
expense is as follows (in thousands):| Fiscal Year Ended August 31,
|     ||-----------------------------------------------------|-----|| 2020
............................................................................. $ 54,165 ||
2021 ............................................................................. 43,780
|| 2022 .............................................................................
28,291 || 2023
............................................................................. 25,877 ||
2024 ............................................................................. 10,976
|| Thereafter .........................................................................
43,174 || **Total
.........................................................................** $206,263 |7.
Accrued ExpensesAccrued expenses consist of the following (in thousands):|
| August 31, 2019 | August 31, 2018
||-------------------------|-----------------|-----------------|| Contract liabilities
| $ 511,329       | —               || Deferred income         | —               | 691,365
|| Accrued compensation    | 600,907         | 570,400         || and employee benefits  |
|                 || Obligation             | 475,251         | —               ||
associated with        |                 |                 || securitization         |
|                 || programs               |                 |                 || Other
accrued expenses | 1,402,657       | 1,000,979       || **Accrued expenses**    |
$2,990,144      | $2,262,744      |8. Notes Payable and Long-Term DebtNotes payable and
long-term debt outstanding as of August 31, 2019 and 2018 are summarized below (in
thousands):|                        | August 31, 2019 | August 31, 2018
||------------------------|-----------------|-----------------|| 5.625% Senior Notes    |
398,886         | 397,995         || (1)(2)                 | Dec 15, 2020    |
|| 4.700% Senior Notes    | 498,004         | 497,350         || (1)(2)                 |
Sep 15, 2022    |                 || 4.900% Senior Notes    | 299,057         | 298,814
|| (1)                    | Jul 14, 2023    |                 || 3.950% Senior Notes    |
494,825         | 494,208         || (1)(2)(3)              | Jan 12, 2028    |
|| Borrowings under      |                 |                 || credit facilities(4)   |
|                 || (5)(6)                 | Nov 8, 2022 and|                 ||
Borrowings under      |                 |                 || loans(4)(5)            |
|                 || (4)                    |                 |                 || Total
notes payable    | 2,496,465       | 2,518,699       || and long-term debt     |
|                 || (1)                    |                 |                 || Less
current           | 375,181         | 25,197          || installments of notes  |
|                 || payable and long-term  |                 |                 || debt
|                 |                 || (2)                    |                 |
|| Total notes payable    | $2,121,284      | $2,493,502      || and long-term debt,    |
|                 || less current install-  |                 |                 || ments
|                 |                 |(1) The notes are carried at the principal amount of
each note, less any unamortized discount and unamortized debt issuance costs.(2) The
Senior Notes are the Company’s senior unsecured obligations and rank equally with all
other existing and future senior unsecured debt obligations.(3) During the fiscal year
ended August 31, 2018, the Company issued $500.0 million of publicly registered 3.950%
Senior Notes due 2028 (the “3.950% Senior Notes”). The net proceeds from the offering were
used.
  \end{Verbatim}
\end{tcolorbox}

\begin{tcolorbox}[
  colback=gray!20,  
  colframe=gray!20, 
  boxrule=0pt,      
  sharp corners    
]
\textbf{Dataset / ID:}\\
convfinqa\_1119 \\ \\
\textbf{Question:}\\
  what was the change in percentage points of data center cost between the years of 2014-13 and 2013-12? \\
  \\
\textbf{Reformulated:}\\
  What was the percentage point decrease in data center cost growth between fiscal 2013-2012 and fiscal 2014-2013 for Adobe Inc.? \\ \\
  \textbf{Context:}\\
\begin{Verbatim}[fontsize=\small,frame=none, baselinestretch=1, breaklines=true]
subscription cost of subscription revenue consists of third-party royalties and expenses
related to operating our network infrastructure , including depreciation expenses and
operating lease payments associated with computer equipment , data center costs , salaries
and related expenses of network operations , implementation , account management and
technical support personnel , amortization of intangible assets and allocated overhead .
we enter into contracts with third-parties for the use of their data center facilities and
our data center costs largely consist of the amounts we pay to these third parties for
rack space , power and similar items . cost of subscription revenue increased due to the
following : % (  % ) change 2014-2013 % (  % ) change 2013-2012 .|  | % (  % )
change2014-2013 | % (  % ) change2013-2012 || --- | --- | --- || data center cost | 10% (
10 % ) | 11% ( 11 % ) || compensation cost and related benefits associated with headcount
| 4 | 5 || depreciation expense | 3 | 3 || royalty cost | 3 | 4 || amortization of
purchased intangibles | 2014 | 4 || various individually insignificant items | 1 | 2014 ||
total change | 21% ( 21 % ) | 27% ( 27 % ) |cost of subscription revenue increased during
fiscal 2014 as compared to fiscal 2013 primarily due to data center costs , compensation
cost and related benefits , deprecation expense , and royalty cost . data center costs
increased as compared with the year-ago period primarily due to higher transaction volumes
in our adobe marketing cloud and creative cloud services . compensation cost and related
benefits increased as compared to the year-ago period primarily due to additional
headcount in fiscal 2014 , including from our acquisition of neolane in the third quarter
of fiscal 2013 . depreciation expense increased as compared to the year-ago period
primarily due to higher capital expenditures in recent periods as we continue to invest in
our network and data center infrastructure to support the growth of our business . royalty
cost increased primarily due to increases in subscriptions and downloads of our saas
offerings . cost of subscription revenue increased during fiscal 2013 as compared to
fiscal 2012 primarily due to increased hosted server costs and amortization of purchased
intangibles . hosted server costs increased primarily due to increases in data center
costs related to higher transaction volumes in our adobe marketing cloud and creative
cloud services , depreciation expense from higher capital expenditures in prior years and
compensation and related benefits driven by additional headcount . amortization of
purchased intangibles increased primarily due to increased amortization of intangible
assets purchased associated with our acquisitions of behance and neolane in fiscal 2013 .
services and support cost of services and support revenue is primarily comprised of
employee-related costs and associated costs incurred to provide consulting services ,
training and product support . cost of services and support revenue increased during
fiscal 2014 as compared to fiscal 2013 primarily due to increases in compensation and
related benefits driven by additional headcount and third-party fees related to training
and consulting services provided to our customers . cost of services and support revenue
increased during fiscal 2013 as compared to fiscal 2012 primarily due to increases in
third-party fees related to training and consulting services provided to our customers and
compensation and related benefits driven by additional headcount , including headcount
from our acquisition of neolane in fiscal 2013. .
  \end{Verbatim}
\end{tcolorbox}

\begin{tcolorbox}[
  colback=gray!20,  
  colframe=gray!20, 
  boxrule=0pt,      
  sharp corners    
]
\textbf{Dataset / ID:}\\
convfinqa\_2966 \\ \\
\textbf{Question:}\\
  what was the value of free cash flow in 2009? \\
  \\
\textbf{Reformulated:}\\
  What was the free cash flow of Union Pacific Corporation in 2009, as calculated from cash provided by operating activities, less cash used in investing activities and dividends paid? \\ \\
  \textbf{Context:}\\
\begin{Verbatim}[fontsize=\small,frame=none, baselinestretch=1, breaklines=true]
2022 asset utilization 2013 in response to economic conditions and lower revenue in 2009 ,
we implemented productivity initiatives to improve efficiency and reduce costs , in
addition to adjusting our resources to reflect lower demand . although varying throughout
the year , our resource reductions included removing from service approximately 26% ( 26 %
) of our road locomotives and 18% ( 18 % ) of our freight car inventory by year end . we
also reduced shift levels at most rail facilities and closed or significantly reduced
operations in 30 of our 114 principal rail yards . these demand-driven resource
adjustments and our productivity initiatives combined to reduce our workforce by 10% ( 10
% ) . 2022 fuel prices 2013 as the economy worsened during the third and fourth quarters
of 2008 , fuel prices dropped dramatically , reaching $ 33.87 per barrel in december 2008
, a near five-year low . throughout 2009 , crude oil prices generally increased , ending
the year around $ 80 per barrel . overall , our average fuel price decreased by 44% ( 44 %
) in 2009 , reducing operating expenses by $ 1.3 billion compared to 2008 . we also
reduced our consumption rate by 4% ( 4 % ) during the year , saving approximately 40
million gallons of fuel . the use of newer , more fuel efficient locomotives ; increased
use of distributed locomotive power ; fuel conservation programs ; and improved network
operations and asset utilization all contributed to this improvement . 2022 free cash flow
2013 cash generated by operating activities totaled $ 3.2 billion , yielding free cash
flow of $ 515 million in 2009 . free cash flow is defined as cash provided by operating
activities , less cash used in investing activities and dividends paid . free cash flow is
not considered a financial measure under accounting principles generally accepted in the
united states ( gaap ) by sec regulation g and item 10 of sec regulation s-k . we believe
free cash flow is important in evaluating our financial performance and measures our
ability to generate cash without additional external financings . free cash flow should be
considered in addition to , rather than as a substitute for , cash provided by operating
activities . the following table reconciles cash provided by operating activities ( gaap
measure ) to free cash flow ( non-gaap measure ) : millions of dollars 2009 2008 2007 .|
millions of dollars | 2009 | 2008 | 2007 || --- | --- | --- | --- || cash provided by
operating activities | $ 3234 | $ 4070 | $ 3277 || cash used in investing activities |
-2175 ( 2175 ) | -2764 ( 2764 ) | -2426 ( 2426 ) || dividends paid | -544 ( 544 ) | -481 (
481 ) | -364 ( 364 ) || free cash flow | $ 515 | $ 825 | $ 487 |2010 outlook 2022 safety
2013 operating a safe railroad benefits our employees , our customers , our shareholders ,
and the public . we will continue using a multi-faceted approach to safety , utilizing
technology , risk assessment , quality control , and training , and by engaging our
employees . we will continue implementing total safety culture ( tsc ) throughout our
operations . tsc is designed to establish , maintain , reinforce , and promote safe
practices among co-workers . this process allows us to identify and implement best
practices for employee and operational safety . reducing grade-crossing incidents is a
critical aspect of our safety programs , and we will continue our efforts to maintain ,
upgrade , and close crossings ; install video cameras on locomotives ; and educate the
public about crossing safety through our own programs , various industry programs , and
other activities . 2022 transportation plan 2013 to build upon our success in recent years
, we will continue evaluating traffic flows and network logistic patterns , which can be
quite dynamic from year-to-year , to identify additional opportunities to simplify
operations , remove network variability and improve network efficiency and asset
utilization . we plan to adjust manpower and our locomotive and rail car fleets to .
  \end{Verbatim}
\end{tcolorbox}

\newpage
\section{Data Preparation}\label{apx:data-prep}

\paragraph{FinQA.}
The FinQA dataset is based on human-annotated questions about documents from FinTabNet, a large corpus of PDF files containing annual reports of S\&P 500 companies. 
In addition to existing data, company-specific information such as founding year, sector, and report year was added. 
Since the answers consisted either of formulas or numerical values, all formulas were parsed and converted into numerical values, as discrepancies between formulas and their numerical solutions were observed. Moreover, approximately 150 yes/no questions were normalized by converting their answers to 0 and 1, respectively.

\paragraph{ConvFinQA.}
The ConvFinQA dataset is also based on FinTabNet and was enriched with additional metadata. Similar to FinQA, answers were standardized by converting formulas and numeric responses into a uniform format. To reduce task complexity and eliminate potential confounding factors, only the first question from each conversation was included. This reduced the dataset size from 14,115 to 3,458 QA pairs.


\paragraph{TAT-DQA.}
TAT-DQA is an independent dataset based on publicly available financial reports. The original dataset included four answer types: Span, Multi-span, Arithmetic, and Count. To ensure consistency with other datasets focused solely on numerical reasoning and to maintain uniform evaluation prompts, Multi-span questions were removed. Additionally, Span answers were normalized by removing symbols such as \$ and \%, and converting words like “million” or “billion” into their numeric equivalents. Dates were also reformatted to the US standard. After these filtering steps, the dataset size was reduced from 16,558 to 11,349 QA pairs.

\section{Reformat Prompt}\label{apx:reformat_prompt}
The prompt for reformulating the questions to be context-independent is given in Figure~\ref{fig:reformulation-prompt}
\begin{figure}[!ht]
    \centering
    \small
    \begin{tcolorbox}[
        colback=white,
        colframe=black,
        arc=0pt,
        boxrule=0.5pt,
        left=10pt,
        right=10pt,
        top=5pt,
        bottom=5pt,
        width=\linewidth
    ]
    \begin{verbatim}
## System Prompt
You are a financial education assistant. Your task is to **rephrase a question** based on a specific
table from a financial document. The goal is to ensure that the question:
- Refers to details that **only make sense in this specific context**
- **Does not use generic phrases** like “based on the data above” or “according to the table”
- Is **not answerable** with any other financial document or context
- Keeps the **original answer correct**
- Sounds natural, precise, and unambiguous
- Try to cut of unnecessary words and phrases
You will also be provided with **metadata** from the document (e.g., company name, report
title, year, section).
Use this metadata to ground the question further in context.
The explanation must:
- Describe the **reasoning steps** required to reach the answer
- Refer to **specific values, labels, rows, or relationships** in the table
- Show that the answer is uniquely valid for this table and **tied to the metadata/context**
### Output Format:
Question:
Answer:
Explanation:       
    \end{verbatim}
    \end{tcolorbox}
    \captionsetup{font=small}
    \caption[Prompt template for the reformulation step.]{System prompt to reformulate the questions.}
    \label{fig:reformulation-prompt}
\end{figure}

\clearpage
\section{Annotation Tool}\label{apx:annotation_tool}
\begin{figure}[!h]
    \centering
    \includegraphics[width=1\linewidth]{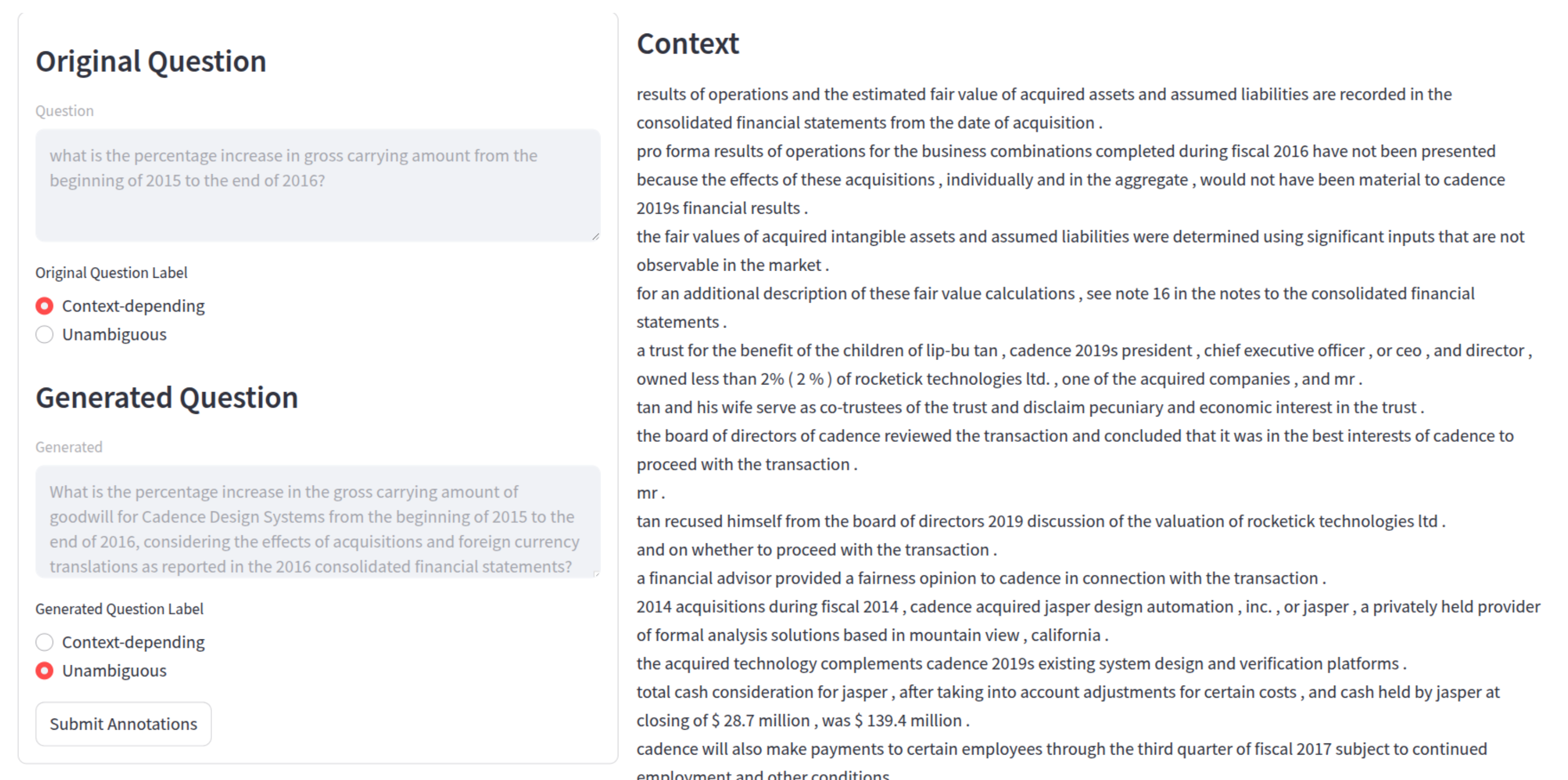}
    \caption{Annotation tool for labeling reformulated questions.}
    \label{fig:annotation-tool}
\end{figure}

The annotations by financial experts were performed with a simple web tool shown in Figure~\ref{fig:annotation-tool}. For each question, the annotator can see the original question, the reformulated question, and the context as given in the dataset. 
\review{The annotators were guided by the following explanations. Annotation Guide: Label the question as 'Context-depending' if the answer depends on the context and can be answered in another context with another true answer, otherwise, label it as 'Unambiguous', when there is only one true answer.}

\newpage
\section{\review{Annotation Samples for Disagreement}}\label{apx:annotation_samples}
\review{
The following six examples illustrate the cases where the commentators disagreed and show where they disagreed. In addition, less than 10\% of the examples were commented on differently.
}
\begin{tcolorbox}[
  colback=gray!20,  
  colframe=gray!20, 
  boxrule=0pt,      
  sharp corners    
]
\textbf{ConvFinQA}\\
\\
\textbf{Original (convfinqa\_10477):}\\
  what was the investment on the alcoainc. in 2014? \\
  \\
\textbf{Reformulated (convfinqa\_5653):}\\
  What was the goodwill balance for Cadence Design Systems as of December 30, 2017, following the business combinations and foreign currency translations during fiscal 2017? \\

\textbf{FinQA}\\
\\
\textbf{Original (train\_finqa1426):}\\
  as of december 312016 what was the ratio of the approximate number of residential vehicles to the large-container industrial? \\
  \\
\textbf{Reformulated (train\_finqa1183):}\\
 What was the percent change in Entergy's net revenue from 2013 to 2014, as reported in the 2015 financial discussion and analysis for Entergy Corporation and Subsidiaries? \\
  
\textbf{TAT-DQA}\\
\\
\textbf{Original (788a22ceb71d2db8786f136e6dd1eed0):}\\
  What was the total value of the changes in principal on the issuance of 2024 Notes, 2026 Notes, 2027 Notes, 2029 Notes, and 2030 Notes? \\
  \\
\textbf{Reformulated (9636d16b010a57a424ab8c02d0f9e46b):}\\
  What percentage of the Australian Prime Storage Fund did National Storage REIT own as at 30 June 2018? \\
\end{tcolorbox}

\newpage
\section{Retrieval Template}\label{apx:retrieval_prompt}
The prompt used to encode the question in the retrieval step is given in Figure~\ref{fig:retrieval-prompt}

\begin{figure}[!ht]
    \centering
    \small
    \begin{tcolorbox}[
        colback=white,
        colframe=black,
        arc=0pt,
        boxrule=0.5pt,
        left=10pt,
        right=10pt,
        top=5pt,
        bottom=5pt,
        width=\linewidth
    ]
    \begin{verbatim}
    Given a question about a company, retrieve relevant passages that answer the query.
    Question:{question}
    \end{verbatim}
    \end{tcolorbox}
    \captionsetup{font=small}
    \caption[Prompt template for the retrieval step.]{System prompt for the retrieval step.}
    \label{fig:retrieval-prompt}
\end{figure}

\newpage
\section{System Prompt for Generation}\label{epx:evaluation_prompt}
We use the same prompt for generating answers (the Generation step in RAG) for all methods we compared. The generation prompt is given in Figure~\ref{fig:generation-prompt-1}-\ref{fig:generation-prompt-3}.
\begin{figure}[!ht]
    \centering
    \small
    \begin{tcolorbox}[
        colback=white,
        colframe=black,
        arc=0pt,
        boxrule=0.5pt,
        left=10pt,
        right=10pt,
        top=5pt,
        bottom=5pt,
        width=\linewidth
    ]
    \begin{verbatim}
YOU ARE A FINANCIAL REASONING EXPERT TRAINED TO ANALYZE A QUESTION AND ITS ASSOCIATED CONTEXT
IN A SINGLE PASS.

  YOUR TASK IS TO:
  - INTERNALLY: READ the question and accompanying financial table/text
    1. UNDERSTAND what the question is asking
    2. IDENTIFY numeric values from the context
    3. CONSTRUCT a valid mathematical FORMULA using a strict symbolic syntax
    4. EVALUATE the formula if it contains only constants
  - FINALLY: OUTPUT one JSON object that includes reasoning, the formula, and the computed result

  THERE IS ONLY ONE INPUT AND ONE OUTPUT. DO ALL THINKING INTERNALLY.
  ---
  FORMULA SYNTAX RULES:

  A formula is either:
  - A number (e.g., 7, 3.14)
  - One of the following symbolic operations, each with exactly two arguments:
    - add(f1, f2)
    - subtract(f1, f2)
    - multiply(f1, f2)
    - divide(f1, f2)
    - exp(f1, f2)
    - greater(f1, f2)

  Nesting is allowed. All values must come from the provided context.
  ---
  PERCENTAGE HANDLING RULES:

  - IF the question asks for a **percentage**, you MUST:
    - REPRESENT the result in the `final_formula` as a **decimal between 0 and 1**
    - COMPUTE the actual percentage internally using divide(part, total)
    - DO NOT multiply by 100 — keep `computed_formula` also between 0 and 1
  - IF a percentage is given in the context (e.g., "12.5%"):
    - CONVERT it to a decimal using divide(12.5, 100) **before using it in a formula**
  - EVEN IF the question says “how much percentage...”, your output stays in **0 to 1 scale**
    - Example: A 12.5% result = "computed_formula": "0.125"
  ---
  OUTPUT FORMAT:
  {
    "reasoning_steps": ["<short bullet 1>", "<short bullet 2>", "..."],
    "final_formula": "<valid formula or 'None'>",
    "computed_formula": "<decimal result as string or 'N/A'>"
  }
  ---
  EXAMPLES:
  EXAMPLE 1 (compute percentage from raw values):

  Input Question:
  What percentage of restricted shares is set to vest after 2021?

  Input Context:
  | Year         | Vesting Count |
  |--------------|----------------|
  | 2021         | 199850         |
  | thereafter   | 110494         |
  | total        | 9038137        |
\end{verbatim}
    \end{tcolorbox}
    \captionsetup{font=small}
    \caption[Prompt template for the generation step.]{System prompt to answer the questions (1/3).}
    \label{fig:generation-prompt-1}
\end{figure}

\begin{figure}[!ht]
    \centering
    \small
    \begin{tcolorbox}[
        colback=white,
        colframe=black,
        arc=0pt,
        boxrule=0.5pt,
        left=10pt,
        right=10pt,
        top=5pt,
        bottom=5pt,
        width=\linewidth
    ]
    \begin{verbatim}
  Output:
  {
    "reasoning_steps": [
      "Located total outstanding restricted shares = 9038137",
      "Found restricted shares vesting after 2021 = 110494",
      "Computed percentage = divide(110494, 9038137)"
    ],
    "final_formula": "divide(110494, 9038137)",
    "computed_formula": "0.01222458878059346"
  }

  ---

  EXAMPLE 2 (compute profit margin — also a percentage):

  Input Question:
  What was the profit margin for 2022?

  Input Context:
  | Year | Revenue   | Net Income |
  |------|-----------|------------|
  | 2022 | 5000000   | 750000     |

  Output:
  {
    "reasoning_steps": [
      "Identified revenue for 2022 = 5000000",
      "Identified net income for 2022 = 750000",
      "Computed profit margin = divide(750000, 5000000)"
    ],
    "final_formula": "divide(750000, 5000000)",
    "computed_formula": "0.15"
  }

  ---

  EXAMPLE 3 (must compute % even if context contains a % value):

  Input Question:
  How much percentage of revenue was allocated to R&D in 2022?

  Input Context:
  | Category      | Amount ($)  |
  |---------------|-------------|
  | Revenue       | 5000000     |
  | R&D Expense   | 625000      |

  Output:
  {
    "reasoning_steps": [
      "Found R&D expense = 625000 and revenue = 5000000",
      "Computed R&D percentage as decimal = divide(625000, 5000000)"
    ],
    "final_formula": "divide(625000, 5000000)",
    "computed_formula": "0.125"
  }

  ---
  \end{verbatim}
    \end{tcolorbox}
    \captionsetup{font=small}
    \caption[Prompt template for the generation step.]{System prompt to answer the questions (2/3).}
    \label{fig:generation-prompt-2}
\end{figure}

\begin{figure}[!ht]
    \centering
    \small
    \begin{tcolorbox}[
        colback=white,
        colframe=black,
        arc=0pt,
        boxrule=0.5pt,
        left=10pt,
        right=10pt,
        top=5pt,
        bottom=5pt,
        width=\linewidth
    ]
    \begin{verbatim}

  UNCLEAR DATA EXAMPLE:

  Input Question:
  What is the average interest coverage ratio?

  Input Context:
  No interest expense or earnings values provided.

  Output:
  {
    "reasoning_steps": [],
    "final_formula": "None",
    "computed_formula": "N/A"
  }

  ---

  STRICT RULES (DO NOT VIOLATE):

  - DO NOT include %, $, €, "million", or any other unit
  - DO NOT guess values or invent data
  - DO NOT return text, markdown, or extra formatting
  - DO NOT multiply by 100 — all percentages must remain in 0–1 decimal form
  - DO NOT use invalid function names or wrong number of arguments
  - DO NOT return “answer”: keys — use only final_formula and computed_formula
  - DO NOT include any formulas or operators in the computed_formula
  - IF a % is provided in the context, convert it to a decimal with divide(X, 100) if needed
      
\end{verbatim}
    \end{tcolorbox}
    \captionsetup{font=small}
    \caption[Prompt template for the generation step.]{System prompt to answer the questions (3/3).}
    \label{fig:generation-prompt-3}
\end{figure}

\clearpage
\section{HyDE Prompt}\label{apx:hyde_prompt}
The prompt used to generate hypothetical documents for the HyDE method is given in Figure~\ref{fig:hyde-prompt}
\begin{figure}[!ht]
    \centering
    \small
    \begin{tcolorbox}[
        colback=white,
        colframe=black,
        arc=0pt,
        boxrule=0.5pt,
        left=10pt,
        right=10pt,
        top=5pt,
        bottom=5pt,
        width=\linewidth
    ]
    \begin{verbatim}
You are a financial analyst. Given a financial question, generate a detailed and realistic 
hypothetical financial document using typical language and structure found in financial reports and 
documents. 
Your answer may include plausible numerical values, trends, and terminology, as if it came from an 
actual financial report. 
The goal is to produce a text that matches the type of content found in financial documents containing 
both text and tables, to aid dense retrieval.
    \end{verbatim}
    \end{tcolorbox}
    \captionsetup{font=small}
    \caption[Prompt for the HyDE method.]{Prompt for the HyDE method.}
    \label{fig:hyde-prompt}
\end{figure}

\section{Summarizing Prompt}\label{apx:summarize_prompt}
The prompt used to generate summarizations for the \emph{Summarization} and \emph{SumContext} methods is given in Figure~\ref{fig:summarization-prompt}.
\begin{figure}[!ht]
    \centering
    \small
    \begin{tcolorbox}[
        colback=white,
        colframe=black,
        arc=0pt,
        boxrule=0.5pt,
        left=10pt,
        right=10pt,
        top=5pt,
        bottom=5pt,
        width=\linewidth
    ]
    \begin{verbatim}
You are a helpful assistant. Your task is to summarize the context text that the user provides 
for better performance in a RAG system.
Pay special attention to all the numerical information, especially those contained in tables.
The summary does not necessarily have to contain all the numerical information, but from 
reading the summary, one should be able to tell what information are contained in the text.
When you receive the context text from the user, ONLY output the summarized text WITHOUT any 
extra reasoning or prefix / postfix text.
    \end{verbatim}
    \end{tcolorbox}
    \captionsetup{font=small}
    \caption[Summarization prompt.]{Summarization prompt.}
    \label{fig:summarization-prompt}
\end{figure}

\newpage

\section{Retrieval Models Source}\label{apx:retrieval_models}
\begin{table}[!h]
\centering
\begin{tabular}{lll}
\toprule
\textbf{Model} & \textbf{Size} & \textbf{Source} \\
\midrule
Stella-EN-1.5B & 1B & \href{https://huggingface.co/NovaSearch/stella_en_1.5B_v5}{NovaSearch/stella\_en\_1.5B\_v5} \\
GTE-Qwen2 1.5B Instruct & 1B & \href{https://huggingface.co/Alibaba-NLP/gte-Qwen2-1.5B-instruct}{Alibaba-NLP/gte-Qwen2-1.5B-instruct} \\
Multilingual E5-Instruct & 560M & \href{https://huggingface.co/intfloat/multilingual-e5-large-instruct}{intfloat/multilingual-e5-large-instruct} \\
Gemini: Text-Embedding-004 & unknown & \href{https://developers.googleblog.com/en/gemini-embedding-text-model-now-available-gemini-api/}{Google Gemini API} \\
OpenAI: Text-Embedding-3 Large & unknown & \href{https://platform.openai.com/docs/guides/embeddings}{OpenAI API Documentation} \\
\bottomrule
\end{tabular}
\caption{Model sizes and sources of evaluated embedding models.}
\label{tab:embedding_sources}
\end{table}

\section{\review{Error Analysis}}\label{apx:error_categories}
\review{
To better understand the model's failure cases, we conducted a manual error analysis on the Oracle-Context setting where the LLaMA 3.3 70B model was used. On average, the model answered 72.7\% of the questions correctly across all subsets. We define the remaining 27.3\% of questions as \textit{error cases}. From these, we randomly sampled 25\% to reduce annotation effort, resulting in a total of 1,583 examples for manual inspection.
}
\review{
To derive meaningful error categories, we began by annotating a small subset of 20 examples from each data split freely. This exploratory step allowed us to identify recurring patterns in the model's failure modes. Based on this qualitative analysis, we established a set of consistent error categories, which are summarized in Table~\ref{tab:error-categories}. Many of the observed errors were systematic and repeated across examples, indicating that our sampled subset provides a representative estimate of the broader error distribution.
}

 \begin{table}[h]
\centering
\small
\begin{tabular}{p{4cm} p{9.5cm}}
\toprule
\textbf{Category} & \textbf{Description and Example} \\
\midrule
\textbf{Miscalculation} & Basic arithmetic mistake (e.g., sum, difference, average). \newline \textit{Example:} subtract(196545, 176675) = 19870, but model returned 19670. \\
\addlinespace
\textbf{Parsing error} & Incorrect extraction of values from table (wrong row/column). \newline \textit{Example:} Summed wrong entries or picked incorrect column values. \\
\addlinespace
\textbf{Over-reasoning} & Performed unnecessary computation instead of direct lookup. \newline \textit{Example:} Answer in plain text, but model tried to compute. \\
\addlinespace
\textbf{Wrong Reformulated Question} & Reformulation subtly changed the metric. \newline \textit{Example:} Original asks for \textit{sum}, reformulation asks for \textit{average}. \\
\addlinespace
\textbf{Wrong Seed Question} & Original query in seed dataset is unanswerable. \newline \textit{Example:} Asked for 2016/17 data when table ends at 2015. \\
\addlinespace
\textbf{Other} & Cases where the answer was \textit{NA}, JSON was parsed incorrectly, or other unclear issues. \newline \textit{Example:} Empty answer, malformed input, or ambiguous logic. \\
\bottomrule
\end{tabular}
\caption{Error categories for model failures with updated labels.}
\label{tab:error-categories}
\end{table}

\end{document}